\documentclass[letterpaper, 10 pt, conference]{ieeeconf}  
\usepackage{comment}
\usepackage{graphics}
\usepackage{amsmath}
\usepackage{amssymb}
\usepackage{amsfonts}
\usepackage{cite}
\usepackage{adjustbox}
\usepackage{tikz}
\usetikzlibrary{positioning,arrows,shapes,chains,fit,scopes}
\usepackage{cite}
\usepackage{upgreek}
\usepackage{textcomp,float}
\usepackage{times} 
\usepackage{euscript}
\usepackage{color}
\usepackage{ccaption}
\usepackage{listings}
\usepackage{longtable}
\usepackage{multirow}
\usepackage{epstopdf}

\usepackage{makecell}
\usepackage{subfigure}
\usepackage{mathtools}
\usepackage{array}
\IEEEoverridecommandlockouts                              
\usepackage{booktabs}   
\usepackage{caption}                                                    
\usepackage{dsfont}

\usepackage{calc}

\newtheorem{Theorem}{\bf \text{Theorem}}
\newtheorem{Definition}{\bf \text{Definition}}
\newtheorem{Proposition}{\bf \text{Proposition}}

\newtheorem{corollary}[Theorem]{\bf \text{Corollary}}
\newtheorem{Lemma}[Theorem]{\bf \text{Lemma}}
\newtheorem{Remark}{\bf \text{Remark}}

\newtheorem{appLemma}{\bf \text{Lemma}}[section]

\let\Sigma\varSigma
\let\Gamma\varGamma

\makeatletter

\newcommand{\Rmnum}[1]{\expandafter \@slowromancap\romannumeral #1@}
\makeatother

\DeclareMathOperator*\argmin{arg\,min}

\DeclareFontFamily{U}{mathx}{}
\DeclareFontShape{U}{mathx}{m}{n}{<-> mathx10}{}
\DeclareSymbolFont{mathx}{U}{mathx}{m}{n}
\DeclareMathAccent{\widehat}{0}{mathx}{"70}
\DeclareMathAccent{\widecheck}{0}{mathx}{"71}

\title{\LARGE \bf
	Characterization of Stable RKHSs with  Sobolev Spaces: from Change of Variables to  Modulation of Amplitude
}

\author{Xiaozhu Fang, Junpeng Zhang, and Tianshi Chen 
	\thanks{This work was supported by NSFC, China under Grant 62273287 and the Shenzhen Science and Technology Innovation Council, China under Grant JCYJ20220530143418040, and the Science and Technology Ph.D. Research Startup Project under Grant No. SZIIT2025KJ060. (Corresponding author: Tianshi Chen.)}
	\thanks{Xiaozhu Fang and Tianshi Chen are with School of Data Science and Shenzhen Research Institute of Big Data, The Chinese University of Hong Kong, Shenzhen, 518172 China (email: 220019189@link.cuhk.edu.cn, tschen@cuhk.edu.cn). Junpeng Zhang is with the School of Artificial Intelligence, Shenzhen University of Information Technology,  Shenzhen 518172, China (jpzhang@suit-sz.edu.cn). }	
}

\begin{document}

	\maketitle
	\thispagestyle{empty}
	\pagestyle{empty}

	\begin{abstract}
	Stable reproducing kernel Hilbert spaces (RKHSs) are employed as the hypothesis space in kernel-based regularized system identification. However, the characterization of stable RKHSs remains challenging because the RKHS norm may not have an explicit expression.  In this work, we  explore the explicit RKHS norm of widely used stable RKHSs,  where an existing  route is revisited and a new route via amplitude modulation is developed.  The existing route is based on the change of variables applied to the spline kernel, and the new route uses modulation of amplitude applied to Sobolev-related stationary kernels. Both routes yield the RKHS norm through an isometric isomorphism, and the norm's expression shows  which decay and smoothness properties are penalized, and how the hyperparameters change that penalty. 		
	\end{abstract}
	
	\section{Introduction}
	The theory of reproducing kernel Hilbert spaces (RKHSs) has attracted considerable attention in the kernel-based regularization method (or simply the kernel method),  which has emerged as a new paradigm in system identification \cite{PDCDL14, CP19, LCM20}.
	One of the key challenges in the kernel method is how to determine the kernel or, equivalently, its uniquely associated RKHS.  The RKHS serves as the hypothesis space in the estimation problem \cite{CS02}, and its determination involves not only the incorporation of prior knowledge about the function to be estimated but also the necessity of ensuring the boundedness of the linear functionals for the representer theorem to hold \cite{Wahba90,DS12}. This has led to growing interest in stable RKHSs whose functions are all (absolutely) integrable,  see, e.g., \cite{ Dinuzzo12,Pillonetto18, BP2020mathematical}, and their associated kernels are referred to as stable kernels. Over the past decade,  many stable kernels have been proposed \cite{PN10a, COL12, CHEN18, ZC18}, such as the stable spline (SS) kernel and the diagonal correlated (DC) kernel.  Most of these stable kernels embed prior knowledge about the stability and smoothness of impulse responses, and they also secure the boundedness of linear functionals \cite{KS23existence, PB2024, FXC26}. 

	The characterization of stable RKHSs  is important for understanding the hypothesis space used in estimation. 
The characterization of a stable RKHS can be achieved by finding the explicit expression for its norm. Given this expression, we are able to understand the membership of the RKHS and the role of the kernel's hyperparameters in the norm.  However, the RKHS norm may not have an explicit expression, and  the characterization of many stable RKHSs  remains vague. 
	
	Fortunately, for some widely used stable RKHSs  in \cite{PN10a, COL12, CHEN18, ZC18}, we can relate them to the Sobolev spaces that possess explicit norm expressions. More specifically,  an isometric isomorphism is established from a Sobolev space onto the targeted stable RKHS, thereby enabling their characterization. 
	The isometric isomorphism  can be established by an existing route in  \cite{PCN10,BP13, Chen19}, where the  isometric isomorphism was established by applying the change of variables (CoV) to the spline kernels or generalized spline kernels, and then the RKHS norms for the SS kernel and DC kernel were derived.
	
	 In this work, we propose a new route, where the isometric isomorphism is established  by applying the modulation of amplitude (MoA) to the Sobolev-related stationary kernels, e.g.,  the Laplace kernel \cite{BT04, SS16}. The new route provides the characterization for a broader class of stable kernels with the  amplitude modulated local stationary  (AMLS) structure, including the SS kernel and DC kernel as special cases. Although both CoV and MoA can recover the SS kernel and DC kernel, MoA allows other amplitude envelopes, and its	 direct norm expression relates the kernel’s hyperparameters to the weighting coefficients in the norm of Sobolev spaces.

	The remainder of this paper is organized as follows. In Section \ref{se:problem}, the estimation problem and RKHS are briefly reviewed. In Section \ref{se:sobolev_space}, Sobolev spaces are introduced. In Section \ref{se:main_result_rkhs_sobolev}, the main results are presented by first revisiting the existing route CoV and then proposing the new route MoA. The proofs of the key results are provided in the  appendix, and the remaining proofs can be found in the journal version  of this paper. . 	
	
	\subsubsection*{Notation} 
	$\mathbb{C}$, $\mathbb{R}$, $\mathbb{R}_+$, $\mathbb{N}$ denote the sets of complex, real,  positive real, and natural numbers, respectively.    
	$\mathds{1}_{\mathcal{X}}(t)$ denotes the indicator function that equals 1 if $t\in \mathcal{X}$ and 0 otherwise.
	$\mathcal{I}_o$ denotes an open interval (connected set) of $\mathbb{R}$ with the Lebesgue measure, possibly unbounded. Some spaces of functions $g:\mathcal{I}_o\rightarrow \mathbb{R}$ are denoted as follows:
 $\mathcal{L}_p(\mathcal{I}_o)$ denotes the space of Lebesgue measurable functions for which the 
		$p$-th power of the absolute value is (absolutely) integrable; $\mathcal{L}_{p,\text{loc}}(\mathcal{I}_o)$ denotes the space of locally $p$-integrable functions;
		$\mathcal{C}_b(\mathcal{I}_o)$  denotes the space of continuous and bounded functions; 	$\mathcal{C}^\infty(\mathcal{I}_o)$  denotes the space of infinitely differentiable functions;  
		$\mathcal{C}_c^\infty(\mathcal{I}_o)$  denotes the space of the so-called test functions, i.e., the infinitely differentiable functions with compact support on $\mathcal{I}_o$;
		$W^{m,p}(\mathcal{I}_o)$  denotes  Sobolev spaces in \eqref{eq:def_sobolev_space}; 	$W^{m,p}_{loc}(\mathcal{I}_o)$  is defined analogously to  $\mathcal{L}_{p,\text{loc}}(\mathcal{I}_o)$. 
	\section{Background and Problem Statement}\label{se:problem}
	
	\subsection{RKHS: a Hypothesis Space in Function Estimation}
	Let $\mathcal{H}_K$ be a reproducing kernel Hilbert space (RKHS) of functions $g:\mathcal{I}_o\rightarrow \mathbb{R}$  associated  with a (positive semi-definite) kernel $K: \mathcal{I}_o\times \mathcal{I}_o\rightarrow \mathbb{R}$. It is worth noting that the kernel $K$ uniquely determines an RKHS $\mathcal{H}_K$ \cite{Aronszajn50}.  More details about kernels and RKHSs can be found in \cite{SC08, BT04}.
	
	Consider the estimation of $g$ by minimizing the kernel-based regularized least squares  (RLS) in the form of 
	\begin{align}\label{eq:rls_rkhs}
		\hat{g} = \argmin\limits_{g\in \mathcal{H}_K} \sum\limits_{i=0}^{N-1}|y_i - L_i[g]|^2 + \gamma\| g\|_{\mathcal{H}_K}^2,
	\end{align}
	where  $\{y_i\}_{i=0}^{N-1}$ are the given observations as the output data, $L_i: \mathcal{H}_K\rightarrow \mathbb{R}$ are the given bounded linear functionals, and $\gamma>0$ is the regularization parameter. It follows from the  representer theorem \cite{Wahba90, SHS01, MP05} that the optimal solution of \eqref{eq:rls_rkhs} can be expressed by a linear combination of the representers that are associated with $\{L_i\}_{i=0}^{N-1}$. 
	In the RLS estimate \eqref{eq:rls_rkhs}, the RKHS $\mathcal{H}_K$  serves as the hypothesis space and must  be determined prior to the estimation.

	\subsection{Stable RKHSs for System Identification}
	
	Impulse response (IR) estimation is a fundamental  problem in system identification. For 
	continuous-time causal and stable\footnote{The stability in this work is referred to as the bounded-input bounded-output (BIBO) stability with the Lebesgue measure, or, equivalently, the $\mathcal{L}_1$-integrability of the impulse response   \cite[Eq. (2.87)]{Oppenheim97}.  } systems,  $\{L_i\}_{i=0}^{N-1}$ in  \eqref{eq:rls_rkhs} take the form of
	\begin{align*}
		L_i[g]= \int_{\mathbb{R}_+}g(s)u(t_i-s)ds, 
	\end{align*}  
	where  $t_i = i\times T$ with  $T >0$ being the sampling interval, $u: \mathbb{R}\rightarrow \mathbb{R}$ is the given measurable and  bounded input signal, and $g: \mathbb{R}_+\rightarrow \mathbb{R}$ represents the causal IR. 
	For clarity, whenever $\mathcal{I}_o=\mathbb{R}_+$, we rewrite the kernel $K: \mathcal{I}_o\times \mathcal{I}_o\rightarrow \mathbb{R}$ as $\kappa: \mathbb{R}_+\times \mathbb{R}_+\rightarrow \mathbb{R}$ regarding the IR.
	
	Over the past decade, the stable kernels/RKHSs defined below have received increased interest because of their association with bounded-input bounded-output (BIBO) stable systems \cite{ Dinuzzo12,Pillonetto18, BP2020mathematical}.  
	\begin{Definition}[ Stable Kernel/RKHS, \cite{Dinuzzo12, PDCDL14}]\label{de:stale_rkhs}
		A kernel  $\kappa: \mathbb{R}_+\times \mathbb{R}_+\rightarrow \mathbb{R}$ is said to be stable (or the RKHS  $\mathcal{H}_\kappa$ is said to be stable) if all $g\in \mathcal{H}_\kappa$ are  $\mathcal{L}_1(\mathbb{R}_+)$-integrable. 
	\end{Definition}
	Several widely used stable kernels have been proposed in system identification, including the second-order stable spline (SS-2) kernel \cite{PCN10}, diagonal correlated (DC) kernel \cite{COL12}, and tuned correlated (TC) kernel, also known as the first-order stable spline (SS-1) kernel \cite{COL12, PCN10}. These kernels are of the form
	\begin{align}
		&\kappa^{\text{SS-2}}(t,t')\!=\! e^{- \beta(t+t')/2}(\frac{1}{2}e^{-\frac{\beta}{6}|t-t'|} -\frac{1}{6}e^{-\frac{\beta}{2}|t-t'|}),\label{eq:ss_kernel}\\
		&\kappa^{\text{DC}}(t,t')\!=\! e^{-\alpha|t-t'|- \beta(t+t')/2},\label{eq:dc_kernel}\\
		&\kappa^{\text{TC/SS-1}}(t,t')\!=\! e^{- \beta\max(t,t')},\label{eq:tc_kernel}
	\end{align} 
	where $[\alpha, \beta]\in \mathbb{R}_+^2$ are hyperparameters, and the scaling coefficient in front of the kernel is omitted without loss of generality. 
	All the above kernels  \eqref{eq:ss_kernel},  \eqref{eq:dc_kernel}, and \eqref{eq:tc_kernel} can be shown to be stable via \cite[Coro. 4.3]{CVT06}.

	\subsection{A Recap of Sobolev Spaces}\label{se:sobolev_space}
	This subsection is devoted to a review of Sobolev spaces \cite{AF03, Brezis10, Leoni17}. Sobolev spaces are defined by the weak derivative. For clarity, we use different notations for ``derivative'':
	\begin{itemize}
		\item $g^{(n)}$ denotes the $n$-th order classical derivative of $g$;
		\item $D^ng$ denotes the $n$-th order weak derivative of $g$. 
	\end{itemize}
	We now present the definitions of the weak derivative and Sobolev spaces. 
	\begin{Definition}[Weak Derivative]\label{de:weak_derivative}
		Consider two functions $f,g\in \mathcal{L}_{1,\text{loc}}(\mathcal{I}_o)$. If 
		\begin{align*}
			\resizebox{\columnwidth}{!}{$
				\begin{aligned}
					\int_{\mathcal{I}_o} g(t) \phi^{(n)}(t)\, dt \!=\! (-1)^n	\int_{\mathcal{I}_o} f(t)  \phi(t)\, dt, \ \forall \phi\in \mathcal{C}_c^\infty(\mathcal{I}_o),
				\end{aligned}$}
		\end{align*} then $f$ is said to be the $n$-th order weak derivative of $g$, denoted by $f=D^ng$.
	\end{Definition}

	\begin{Definition}[Sobolev space $W^{m,p}$]\label{deF:sobolev_space}
		Let $m$ be a positive integer and $1\leq p \leq \infty$.
		The Sobolev space $W^{m,p}(\mathcal{I}_o)$ is defined as 
		\begin{multline}\label{eq:def_sobolev_space}
			W^{m,p}(\mathcal{I}_o)= \{g: \mathcal{I}_o\rightarrow \mathbb{R}\ |\  \forall n\in\{0,1,\cdots,  m\}, \\ \exists f_n \in \mathcal{L}_p(\mathcal{I}_o) \text{ such that }f_n=D^n g\}. 
		\end{multline}
	\end{Definition}

	The Sobolev space $W^{m,p}(\mathcal{I}_o)$ does not specify the norm.  By default, $W^{m,p}(\mathcal{I}_o)$ for $p<\infty$ is equipped with the norm
	\begin{align}\label{eq:sobolev_norm_general}
		\|g\|_{	W^{m,p}(\mathcal{I}_o)}= \bigg(\sum\limits_{0\leq n\leq m} \| D^n g\|_{\mathcal{L}_p(\mathcal{I}_o)}^p\bigg)^{1/p},
	\end{align}
	and the essential supremum is used for $p=\infty$. 
	With this norm, it can be shown that  $W^{m,p}(\mathcal{I}_o)$ is a Banach space \cite[Thm 3.3]{AF03}, and $W^{m,2}(\mathcal{I}_o)$ is a  Hilbert space \cite[Thm 3.6]{AF03}. Moreover,  $W^{m,2}(\mathcal{I}_o)$  is related to the RKHS, see \cite[Sec. 1.3]{SS16} and \cite[Sec. 7.4]{BT04}, though the equipped norm may be a weighted version of  \eqref{eq:sobolev_norm_general}. 
	
	\begin{Remark}\label{re:abs_cont_sobolev}
		The Sobolev space $W^{m,p}(\mathcal{I}_o)$ can be equivalently characterized using the absolutely continuous (AC) representatives and classical derivatives as
		\begin{multline}\label{eq:def_sobolev_space2}
			W^{m,p}(\mathcal{I}_o)\!=\! \{g:\mathcal{I}_o\rightarrow \mathbb{R}\ |\  g^{(j)} \text{ is AC for } j\!=\!0,\dots,m-1, \\ \text{and } g^{(n)} \in \mathcal{L}_p(\mathcal{I}_o) \text{ for } n=0,\dots,m\}.
		\end{multline}  AC functions have  classic derivatives almost everywhere (a.e.) \cite[Thm. 3.35]{Folland99}.
		The description \eqref{eq:def_sobolev_space2} has been adopted in \cite{Wahba90, PCN10,BP13}, whereas we stick to the description \eqref{eq:def_sobolev_space} for Section \ref{se:moa}. 
	\end{Remark}
	
	\subsection{Problem Statement: Characterization of Stable RKHS}\label{se:pro_state_stable_rkhs}
In this work, we are interested in the characterization of the stable RKHSs. The essence of this characterization is to find the explicit form of the RKHS norm, where this norm not only identifies which functions belong to this RKHS but also which functions have a smaller penalty in the RLS estimate \eqref{eq:rls_rkhs}. Moreover, the explicit expression of the norm shows how the hyperparameters change that penalty. 
	However, the explicit norm is typically unavailable for RKHSs. In what follows, we exploit Sobolev spaces to derive explicit norms for a class of stable RKHSs and provide their characterization.

	\section{Main Results}\label{se:main_result_rkhs_sobolev}
	
	By tracing the development of stable RKHSs, there are two  routes for proposing a stable RKHS or kernel:
	\begin{itemize}
		\item \cite{PN10a} introduced the change of variables (CoV) applied to spline kernels, which led to the class of stable spline (SS) kernels;
		\item  \cite{CHEN18} introduced the modulation of amplitude (MoA) applied to stationary kernels, which led to the class of amplitude modulated local stationary (AMLS) kernels.
	\end{itemize}
	We refer to these as the route  CoV and the  route MoA, respectively. The characterization of a stable RKHS can be established within each route by establishing an isometric isomorphism (i.e., a linear bijective norm-preserving mapping)  from a suitably normed Sobolev space $W^{m,2}(\mathcal{I}_o)$ onto the target stable RKHS.
	
	In what follows, we first revisit the route  CoV in Section \ref{se:cov_revisted} and then consider the  route MoA  in  Section \ref{se:moa}.

	%

	\subsection{Change of Variables for the Spline Kernels, Revisited}\label{se:cov_revisted}

	The spline kernels for learning problems can be found in \cite[Sec. 1.2]{Wahba90}, where the $m$-th order spline (Sp-$m$) is denoted by   $K^{\text{Sp-$m$}}:(0,1)\times  (0,1)\rightarrow \mathbb{R}$ with 
	\begin{align}\label{eq:sp_m_kernel}
		&K^{\text{Sp-$m$}}(s,s')=\int_{0}^{1} G_m(s,u)G_m(s',u)\, du,\\
		&G_m(r,u)= \frac{ (r-u)^{m-1}\mathds{1}_{\mathbb{R}_+}(r-u)}{(m-1)!},\nonumber
	\end{align} 
	and its associated RKHS can be characterized by  
	\begin{align}
		&\mathcal{H}_{K^{\text{Sp-$m$}}}\!=\! \{ q\!\in\! W^{m,2}((0,1))\ | \ D^nq(0^+)\!=\!0,  0\!\leq\! n\!\leq\! m-1\},\nonumber\\
		&\|q\|_{\mathcal{H}_{K^{\text{Sp-$m$}}}}^2=\int_{0}^{1}|D^mq(s)|^2 \, ds. \nonumber
	\end{align}

	We now consider a change of variables (CoV) $s = \varphi(t)$ mapping $t \in \mathbb{R}_+$ to $s \in (0,1)$, as introduced in \cite{PCN10}. For generality, we allow for different choices of $\varphi$. This CoV induces the $m$-th order stable spline (SS-$m$-$\varphi$) kernel:
	\begin{align}\label{eq:ss_m_kernel}
		\kappa^{\text{SS-$m$-$\varphi$}}(t,t')=  K^{\text{Sp-$m$}}(\varphi(t),\varphi(t')),\quad t,t'\in \mathbb{R}_+.
	\end{align}
In fact, we can choose an arbitrary CoV $s = \varphi(t)$, but not all of them guarantee the stability of $\kappa^{\text{SS-$m$-$\varphi$}}$. The following sufficient
condition guarantees stability.
	\begin{Lemma}\label{le:stab_cov}
		If $\varphi:\mathbb{R}_+\rightarrow (0,1)$ is measurable and  $\varphi^{m-1/2}\in \mathcal{L}_1(\mathbb{R}_+)$, then 	$\kappa^{\text{SS-$m$-$\varphi$}}$ is stable. 
	\end{Lemma}

Further, if $\varphi$ is bijective, then we have the characterization of $\mathcal{H}_{\kappa^{\text{SS-$m$-$\varphi$}}}$ below. This result has been well established in the learning literature. 
	\begin{Proposition}[\protect{\cite[Prop. 3.3]{CVTU10}}]\label{pr:lemma_cond_varphi}
		Consider the Sp-$m$ kernel \eqref{eq:sp_m_kernel} and the SS-$m$-$\varphi$ kernel \eqref{eq:ss_m_kernel}  with the CoV $\varphi$.
		If $\varphi$ is bijective, then the linear operator
		\begin{align*}
			\mathcal{H}_{K^{\text{Sp-$m$}}} \ni	(s\mapsto q(s) )\mapsto  (t \mapsto q(\varphi(t)))\in \mathcal{H}_{\kappa^{\text{SS-$m$-$\varphi$}}}
		\end{align*}
		is an isometric isomorphism, and  the RKHS associated with the SS-$m$-$\varphi$ kernel \eqref{eq:ss_m_kernel} can be characterized by
		\begin{align}
			&\mathcal{H}_{\kappa^{\text{SS-$m$-$\varphi$}}}\!= \!\{ g:\mathbb{R}_+\rightarrow \mathbb{R}\ | \  g\circ \varphi^{-1}\in W^{m,2}((0,1)), \nonumber \\ 
			&\qquad D^n (g\circ \varphi^{-1})(0^+)=0, \ 0\leq n\leq m-1\},\label{eq:rkhs_chara_ss-m}\\
			&\|g\|_{\mathcal{H}_{\kappa^{\text{SS-$m$}-\varphi}}}^2=\int_{0}^{1}|D^m (g\circ \varphi^{-1})(s)|^2 \, ds. \label{eq:rkhs_norm_ss-m}
		\end{align} 
	\end{Proposition}
	
	In practice, the computation of $D^m (g\circ \varphi^{-1})$ could be cumbersome. 
	If $\varphi$ is a $\mathcal{C}^m$ diffeomorphism\footnote{Let $\mathcal{I}_1$ and $\mathcal{I}_2$ be open intervals of $\mathbb{R}$.  A mapping $\varphi:\mathcal{I}_1\rightarrow \mathcal{I}_2$ is said to be a $\mathcal{C}^m$ diffeomorphism \cite[p. 74]{Folland99}  if it is bijective, and both $\varphi$ and its inverse $\varphi^{-1}$ are $m$-times continuously differentiable.},  then we can employ \cite[Thm. 3.41]{AF03} on every bounded open interval in $\mathbb{R}_+$, such that			\begin{align*}
	\mathcal{H}_{\kappa^{\text{SS-$m$-$\varphi$}}} \subset W^{m,2}_{loc}(\mathbb{R}_+),
	\end{align*}
	which implies that both $g$ and $g\circ \varphi^{-1}$ in \eqref{eq:rkhs_chara_ss-m} are differentiable a.e. in their respective domains. Then we can apply the chain rule \cite[Thm. 3.44]{Leoni17} to $D^n (g\circ \varphi^{-1})$ in \eqref{eq:rkhs_norm_ss-m}.  For instance, the RKHS norms for the TC/SS-1 kernel \eqref{eq:tc_kernel} and the SS-2 kernel \eqref{eq:ss_kernel}, with $\varphi(t) = e^{-\beta t}$ and   $\varphi(t) = e^{-\beta t/3}$, respectively, can be represented by  \cite{PCN10}:

	\begin{subequations}\label{eq:rkhs_norm_special_cov}
		\begin{align}
			&\|g\|_{\mathcal{H}_{\kappa^{\text{TC/SS-1}}}}^2=\int_{\mathbb{R}_+}|Dg(t)|^2 \frac{e^{\beta t}}{\beta}\, dt,\label{eq:rkhs_norm_tc_cov}\\
			&\|g\|_{\mathcal{H}_{\kappa^{\text{SS-2}}}}^2=\int_{\mathbb{R}_+}|D^2g(t)+ \frac{\beta}{3} Dg(t)|^2 \frac{e^{\beta t}}{(\frac{\beta}{3})^3}\, dt.\label{eq:rkhs_norm_ss_cov}
		\end{align}
		An extension to the generalized first-order spline (GSp-1) kernel \cite{Chen19}  yields the RKHS norm for the DC kernel \eqref{eq:dc_kernel}:
		\begin{align}
			&\|g\|_{\mathcal{H}_{\kappa^{\text{DC}}}}^2=	\frac{1}{2\alpha } \int_{\mathbb{R}_+}| Dg(t)-(\alpha-\frac{\beta}{2}) g(t)|^2  e^{\beta t} dt.\label{eq:rkhs_norm_dc_cov}
		\end{align}
	\end{subequations}

	\subsection{ Modulation of Amplitude for the Laplace Kernel }\label{se:moa}
	
	Stationary kernels have been extensively studied in the machine learning literature \cite{RasmussenW:06}. Several stationary kernels are related to Sobolev spaces, including the Matérn kernel \cite[Eq. (4.14)]{RasmussenW:06} and the kernels discussed in \cite{ThomasAgnan96}. We begin with the Laplace kernel, i.e., the Matérn kernel with order $\nu = 0.5$,  whose associated RKHS is characterized in the following lemma.
	\begin{Lemma}[Laplace Kernel]\label{le:rkhs_norm_laplace_kernel}
		Consider the Laplace kernel 	$\kappa^\text{La}:\mathbb{R}_+\times \mathbb{R}_+\rightarrow \mathbb{R}$ given by
		\begin{align}\label{eq:laplace_kernel}
			\kappa^\text{La}(t,t')=\frac{1}{2\alpha}e^{-\alpha |t-t'|}, \quad \alpha>0.
		\end{align}
		Then, the RKHS associated with the Laplace kernel \eqref{eq:laplace_kernel}  can be characterized by 
		\begin{align}
			&\mathcal{H}_{\kappa^{\text{La}}}= W^{1,2}(\mathbb{R}_+),\label{eq:inclusion_dc_sobolev_w12}\\
			&\|h\|_{\mathcal{H}_{\kappa^\text{La}}}^2 =\int_{\mathbb{R}_+}|Dh(t)|^2 dt+\alpha^2 \int_{\mathbb{R}_+} |h(t)|^2 \,dt + \alpha |h(0^+)|^2,\nonumber
		\end{align}
		where the identity in \eqref{eq:inclusion_dc_sobolev_w12} indicates the same elements but different metrics. 
	\end{Lemma}
	Now we consider the element-wise product of the Laplace kernel \eqref{eq:laplace_kernel} and another rank-1 kernel. The resultant kernel has the structure of the amplitude modulated local stationary (AMLS) kernel \cite{CHEN18}, and thereby we call it the amplitude modulated  Laplace (AMLa) kernel.  Now we establish the isometric isomorphism. 
	\begin{Proposition}[AMLa Kernel]\label{pr:rkhs_norm_aml_kernel}
		Consider the  AMLa kernel $\kappa^{\text{AMLa}}:\mathbb{R}_+\times \mathbb{R}_+\rightarrow \mathbb{R}$ given by 
		\begin{align}\label{eq:kernel_aml}
			\kappa^{\text{AMLa}}(t,t')=  b(t)b(t') \kappa^\text{La}(t,t'). 
		\end{align}
		If  $b\in \mathcal{C}_b(\mathbb{R}_+)$ and $b(t) \neq 0$ for a.e. $t\in \mathbb{R}_+$, then the linear operator 
		\begin{align*}
			\mathcal{H}_{\kappa^\text{La}}\ni	(t\mapsto g(t)) \mapsto (t\mapsto b(t)g(t))\in \mathcal{H}_{\kappa^\text{AMLa}}
		\end{align*}
		is an isometric isomorphism. Consequently, the RKHS associated with the AMLa kernel \eqref{eq:kernel_aml} can be characterized by
		\begin{align}
			&\mathcal{H}_{\kappa^{\text{AMLa}}}= \{  g:\mathbb{R}_+\rightarrow \mathbb{R} \ | \ g=bh,\ h\in W^{1,2}(\mathbb{R}_+) \},  \label{eq:aml_rkhs_incluisive_W_12}  \\
			&\resizebox{\columnwidth}{!}{$
				\begin{aligned}\|g\|_{\mathcal{H}_{	\kappa^{\text{AMLa}}}}^2\!=\! \int_{\mathbb{R}_+}   \bigg| D(\frac{g}{b})(t)\bigg|^2dt 
					\!+\!\alpha^2 \int_{\mathbb{R}_+} \bigg|\frac{g}{b}(t)\bigg|^2dt\!+\!  \alpha \bigg|\frac{g}{b}(0^+)\bigg|^2,\end{aligned}$} \label{eq:rkhs_norm_aml_kernel}
		\end{align}
		where $g/b$ is associated with the unique $h$ in \eqref{eq:aml_rkhs_incluisive_W_12}. 
	\end{Proposition}
	 Stability requires an additional condition on $b$. In particular, $b\in\mathcal{L}_2(\mathbb{R}_+)$ is sufficient, since
		$\|g\|_{\mathcal{L}_1}\leq\|b\|_{\mathcal{L}_2}\|g/b\|_{\mathcal{L}_2}.$
	Moreover, if $b\in W^{1,2}(\mathbb{R}_+)\subset \mathcal{C}_b(\mathbb{R}_+)$ and $b(t) \neq 0$ for a.e. $t\in \mathbb{R}_+$, then we can employ Lemma \ref{le:prod_rule_weak_derivative2} to obtain 
	\begin{align*}
		\mathcal{H}_{\kappa^{\text{AMLa}}}\subset  W^{1,2}(\mathbb{R}_+). 
	\end{align*}
	Then we can apply the  product rule (Lemma \ref{le:prod_rule_weak_derivative2}) to $D(\frac{g}{b})$ in \eqref{eq:rkhs_norm_aml_kernel}. Likewise to  \eqref{eq:rkhs_norm_special_cov}, 
	the results of Proposition \ref{pr:rkhs_norm_aml_kernel}  specialize to those for the DC kernel \eqref{eq:dc_kernel} and TC/SS-1 kernel \eqref{eq:tc_kernel}. 
	\begin{corollary}[DC Kernel]\label{co:rkhs_norm_dc_kernel}
		The DC kernel \eqref{eq:dc_kernel} is a special case of the AMLa kernel \eqref{eq:kernel_aml} with  
		$b(t)=\sqrt{2\alpha }e^{-\frac{\beta}{2} t}$. Then,  the RKHS  associated with the DC kernel \eqref{eq:dc_kernel}   can be characterized by 
		\begin{align}
			&\mathcal{H}_{\kappa^{\text{DC}}}\!=\! \{  g:\mathbb{R}_+\rightarrow \mathbb{R} \ |\ \frac{g}{b}\in W^{1,2}(\mathbb{R}_+)  \}\subset  W^{1,2}(\mathbb{R}_+)\nonumber,\\
			&\|g\|_{\mathcal{H}_{\kappa^{\text{DC}}}}^2 
			=\frac{1}{2\alpha} \int_{\mathbb{R}_+} |Dg(t)|^2 e^{\beta t} dt +\nonumber\\ & \left(\frac{\alpha}{2}\!-\!\frac{\beta^2}{8\alpha}\right) \int_{\mathbb{R}_+} |g(t)|^2 e^{\beta t} dt
			\!+\! \left(\frac{1}{2}\!-\!\frac{\beta}{4\alpha}\right) |g(0^+)|^2,\label{eq:rkhs_norm_dc_kernel}
		\end{align}
		where \eqref{eq:rkhs_norm_dc_kernel} is equivalent to \eqref{eq:rkhs_norm_dc_cov}.
	\end{corollary}
	
	\begin{Remark}
Note that  some expanded coefficients in \eqref{eq:rkhs_norm_dc_kernel} are negative when $\alpha < \beta/2$. The  positivity of the norm can be guaranteed from the unexpanded norm \eqref{eq:rkhs_norm_aml_kernel}, not from
	the signs of these coefficients.
	\end{Remark}
	\begin{corollary}[TC/SS-1 Kernel]\label{co:rkhs_norm_Tc_kernel}
		The TC/SS-1 kernel \eqref{eq:tc_kernel} is a special case of the AMLa kernel \eqref{eq:kernel_aml} with  
		$b(t)=\sqrt{2\alpha }e^{-\frac{\beta}{2} t}$ and  $\alpha = \frac{\beta}{2}$. Then,  the RKHS  associated with the TC/SS-1 kernel \eqref{eq:tc_kernel}  can be characterized by 
		\begin{align*}
			\mathcal{H}_{\kappa^{\text{TC/SS-1}}}\!=\!  \{  g:\mathbb{R}_+\rightarrow \mathbb{R} \ |\ \frac{g}{b}\in W^{1,2}(\mathbb{R}_+)  \}\subset  W^{1,2}(\mathbb{R}_+),     
		\end{align*}
		with norm $\| g\|_{\mathcal{H}_{\kappa^{\text{TC/SS-1}}}}$ given in \eqref{eq:rkhs_norm_tc_cov}. 
	\end{corollary}

	\subsection{Towards Higher-Order Regularity}
	It is natural to question whether the route MoA  can be extended to achieve higher-order regularity, in analogy with the extension from the TC/SS-1 kernel \eqref{eq:tc_kernel} to the SS-2 kernel \eqref{eq:ss_kernel}. To address this question, the key is to identify a higher-order Laplace (La-$m$) kernel $\kappa^{\text{La-$m$}}:\mathbb{R}_+\times \mathbb{R}_+\rightarrow \mathbb{R}$ whose associated RKHS norm takes the form
	\begin{align}\label{eq:rkhs_norm_higher_order_regu}
		\|h\|_{\mathcal{H}_{\kappa^{\text{La-$m$}}}}^2 =  \sum\limits_{i=0}^{m}  p_i  \int_{\mathbb{R}_+}|D^i h(t)|^2\,dt+ \text{BT}, 
	\end{align}
	where $p_i \in \mathbb{R}$ are design coefficients with $p_m = 1$ and the other constraints reported later, and BT denotes boundary terms involving $D^i h(0^+)$ for $i \leq m-1$. Once such a stationary kernel is available, stable RKHSs can be constructed via the amplitude modulated (AMLa-$m$) kernel given by 
	\begin{align}\label{eq:amla_2_kernel}
		\kappa^{\text{AMLa-$m$}}(t,t')=  b(t)b(t') \kappa^\text{La-$m$}(t,t'),
	\end{align} 
	where  $b\!\in\! \mathcal{C}_b\cap \mathcal{L}_2(\mathbb{R}_+)$ and $b(t) \neq 0$ for a.e. $t\in \mathbb{R}_+$. 
	
	Owing to space limitations, we briefly outline the approach for finding such an La-$m$ kernel associated with \eqref{eq:rkhs_norm_higher_order_regu}. A general method is described in \cite[Sec. 2]{ThomasAgnan96}, which involves solving a differential equation. Adapting this method to causal kernels on $\mathbb{R}_+ \times \mathbb{R}_+$ is not hard. It is worth noting that the coefficients $p_i$ cannot be chosen arbitrarily: they must satisfy $1 / \big( \sum_{i=0}^{m} p_i \omega^{2i} \big) > 0$ for all $\omega \in \mathbb{R}$ to ensure the positive definiteness of stationary kernels, in accordance with Bochner's theorem \cite{Bochner34}.

	Following the above idea, we can find the second-order  Laplace (La-2) kernel as follows. 
	\begin{Proposition}[La-2 Kernel]\label{pr:amls_norm_higher_regularity}
		Consider the second-order   Laplace (La-2) kernel $\kappa^{\text{La-2}}:\mathbb{R}_+\times \mathbb{R}_+\rightarrow \mathbb{R}$ given by  
		\begin{align}\label{eq:second_order_stable_amls}
			\kappa^{\text{La-2}}(t,t')= \bigg( \frac{e^{-\alpha_1|t-t'| }}{2\alpha_1(\alpha_2^2-\alpha_1^2) }	-\frac{	e^{-\alpha_2|t-t'| }}{2\alpha_2(\alpha_2^2-\alpha_1^2)}\bigg),
		\end{align}
		where $\alpha_1,\alpha_2>0$ and $\alpha_1\neq\alpha_2$.
		Then,  the RKHS associated with the La-2 kernel \eqref{eq:second_order_stable_amls}  can be characterized by 
		\begin{align}
			&	\mathcal{H}_{\kappa^{\text{La-2}}}=  W^{2,2}(\mathbb{R}_+),    \nonumber \\
			&	\resizebox{\columnwidth}{!}{$
			\begin{aligned}	&  \|h\|_{\mathcal{H}_{\kappa^{\text{La-2}}}}^2 =  \int_{\mathbb{R}_+}|D^2h(t)|^2 dt +(\alpha_1^2+\alpha_2^2)  \int_{\mathbb{R}_+}|Dh(t)|^2 dt\\
			&+\alpha_1^2\alpha_2^2 \int_{\mathbb{R}_+} |h(t)|^2 \,dt + \begin{bmatrix}
				Dh(0^+)\\
				h(0^+)
			\end{bmatrix}^\text{T}\begin{bmatrix}
				\alpha_1 \!+\! \alpha_2& -\alpha_1\alpha_2\\
				-\alpha_1\alpha_2& \alpha_1\alpha_2(	\alpha_1 \!+\! \alpha_2)
			\end{bmatrix}\begin{bmatrix}
				Dh(0^+)\\
				h(0^+)
			\end{bmatrix}.\end{aligned}$}\label{eq:rkhs_norm_la2_kernel} 
		\end{align}
	\end{Proposition}
	\begin{corollary}[SS-2 Kernel]
		The SS-2 kernel \eqref{eq:ss_kernel} is a special case of the AMLa-2 kernel \eqref{eq:amla_2_kernel} with $\kappa^{\text{La-2}}$ in  \eqref{eq:second_order_stable_amls}, 
		$b(t)=(\frac{\beta}{3})^{\frac{3}{2}}   e^{-                             \frac{\beta}{2} t}$,  $\alpha_1=\frac{\beta}{6}$, and $\alpha_2=\frac{\beta}{2}$. Then,  the RKHS  associated with the SS-2 kernel \eqref{eq:ss_kernel}  can be characterized by 
		\begin{align*}
			\mathcal{H}_{\kappa^{\text{SS-2}}}= \{   g:\mathbb{R}_+\rightarrow \mathbb{R} \ | \frac{g}{b}\in   W^{2,2}(\mathbb{R}_+) \}\subset W^{2,2}(\mathbb{R}_+),
		\end{align*}
		with norm $\| g\|_{\mathcal{H}_{\kappa^{\text{SS-2}}}}$ given by  \eqref{eq:rkhs_norm_ss_cov} or \eqref{eq:rkhs_norm_la2_kernel} with $h=g/b$.   
	\end{corollary}

	%

	\section{Conclusion}
	This paper has discussed two Sobolev-space-based routes for characterizing stable reproducing kernel Hilbert spaces (RKHSs). The new characterization via amplitude modulation provides guidance for kernel design and deepens  theoretical understanding of RKHS-based hypothesis spaces that have been influential in  system identification.

	\bibliographystyle{ieeetr}
	\bibliography{database}
	
	\def\thesectiondis{\thesection.}                   
	\def\thesubsectiondis{\thesection.\arabic{subsection}.}          
	\def\thesubsubsectiondis{\thesubsection.\arabic{subsubsection}.}
	
	\setcounter{subsection}{0}
	
	\renewcommand{\thesection}{A}

	\renewcommand{\theequation}{A.\arabic{equation}}
	\setcounter{equation}{0}
	
	\renewcommand{\thesubsection}{\thesection.\arabic{subsection}}
	
	\renewcommand{\thesection}{A}
	\renewcommand{\thesubsection}{A.\arabic{subsection}}

	\section*{Appendix}

	\renewcommand{\thesubsection}{A.\arabic{subsection}}
	
	\setcounter{appLemma}{0}
	\renewcommand{\theappLemma}{A.\arabic{appLemma}}
	\setcounter{appRemark}{0}
	\renewcommand{\theappRemark}{A.\arabic{appRemark}}

	\subsection{Technical Lemmas}\label{se:tech_lemmas}
	\begin{appLemma}[Product Rule, Type 1]\label{le:prod_rule_weak_derivative}
		Consider two functions $g\in \mathcal{L}_{1,\text{loc}}(\mathcal{I}_o)$ and $h\in \mathcal{C}^\infty(\mathcal{I}_o)$. If $Dg\in \mathcal{L}_{1,\text{loc}}(\mathcal{I}_o)$ exists,  then  $D(gh)\in   \mathcal{L}_{1,\text{loc}}(\mathcal{I}_o)$ exists and  
		\begin{align}\label{eq:prod_rule_weak_derivative}
			D(gh)= Dg h+ g h', 
		\end{align}
		where $h'$ denotes the derivative of $h$. 
	\end{appLemma}
	\begin{appLemma}[Product Rule, Type 2, \protect{\cite[Coro. 8.10]{Brezis10}}]\label{le:prod_rule_weak_derivative2}
	Let $1\leq p \leq \infty$.	Consider two functions $g\in W^{1,p}(\mathcal{I}_o)$ and $h\in W^{1,p}(\mathcal{I}_o)$. Then,  $gh\in W^{1,p}(\mathcal{I}_o)$ and 
		\begin{align}\label{eq:prod_rule_weak_derivative2}
			D(gh )= Dgh+ gDh.	\end{align}
	\end{appLemma}

	%
	%

	\subsection{Proof of Lemma \ref{le:stab_cov}}
	The diagonal of the $m$-th order spline kernel satisfies
	\begin{align*}
		K^{\mathrm{Sp}\text{-}m}(s,s)
		&=
		\frac{s^{2m-1}}
		{(2m-1)((m-1)!)^2}.
	\end{align*}
	Hence,
		$\sqrt{
			\kappa^{\mathrm{SS}\text{-}m\text{-}\varphi}(t,t)}
		=
		\varphi(t)^{m-\frac12}/(
		(m-1)!\sqrt{2m-1})$.
	For every
	$g\in\mathcal{H}_{\kappa^{\mathrm{SS}\text{-}m\text{-}\varphi}}$,
	the reproducing property and the Cauchy--Schwarz inequality give
	\begin{align*}
		\resizebox{\columnwidth}{!}{$
			\begin{aligned}
				|g(t)|
				&\leq
				\|g\|_{\mathcal{H}_{\kappa^{\mathrm{SS}\text{-}m\text{-}\varphi}}}
				\sqrt{
					\kappa^{\mathrm{SS}\text{-}m\text{-}\varphi}(t,t)}=\frac{
					\|g\|_{\mathcal{H}_{\kappa^{\mathrm{SS}\text{-}m\text{-}\varphi}}}
				}
				{(m-1)!\sqrt{2m-1}}
				\varphi(t)^{m-\frac12}.\end{aligned}$}
	\end{align*}
	Since $\varphi$ is measurable and
	$\varphi^{m-\frac12}\in\mathcal{L}_1(\mathbb{R}_+)$, integration over
	$\mathbb{R}_+$ gives
	\begin{align*}
		\|g\|_{\mathcal{L}_1(\mathbb{R}_+)}
		\leq
		\frac{
			\|\varphi^{m-\frac12}\|_{\mathcal{L}_1(\mathbb{R}_+)}
		}
		{(m-1)!\sqrt{2m-1}}
		\|g\|_{\mathcal{H}_{\kappa^{\mathrm{SS}\text{-}m\text{-}\varphi}}}
		<\infty,
	\end{align*}
	which proves that
	$\kappa^{\mathrm{SS}\text{-}m\text{-}\varphi}$ is stable.

	\subsection{Proof of Lemma \ref{le:rkhs_norm_laplace_kernel}}
	
	The proof of Lemma \ref{le:rkhs_norm_laplace_kernel} is routine and analogous to the examples in \cite[Sec. 1.3]{SS16}. However, due to the subtle difference at the boundary  $t=0$, we 
	elaborate the  proof below. 
	Let $\mathcal{V}(\mathbb{R}_+)$ denote a normed space by
	\begin{align*}
		&\mathcal{V}(\mathbb{R}_+):= \{h\in W^{1,2}(\mathbb{R}_+) \ |\ \|h\|_{\mathcal{V}(\mathbb{R}_+)} <\infty \},\\
		&\|h\|_{\mathcal{V}(\mathbb{R}_+)}^2 =\int_{\mathbb{R}_+}|Dh(t)|^2 dt+\alpha^2 \int_{\mathbb{R}_+} |h(t)|^2 \,dt + \alpha |h(0^+)|^2.
	\end{align*}

	First, we aim to show that  $\mathcal{V}(\mathbb{R}_+)= W^{1,2}(\mathbb{R}_+)$, i.e., $\mathcal{V}(\mathbb{R}_+)$ has the same functions as $W^{1,2}(\mathbb{R}_+)$. To this end, we only need to show their norms are equivalent, i.e., there exist $0<C_1, C_2<\infty$ such that 
	\begin{align*}
		C_1\|h\|_{W^{1,2}(\mathbb{R}_+)}\leq \|h\|_{\mathcal{V}(\mathbb{R}_+)}\leq C_2\|h\|_{W^{1,2}(\mathbb{R}_+)}. 
	\end{align*}
	The existence of $C_1$ is trivial, and now we show the existence of $C_2$. To this end, we first show $h(0^+)$ is finite. We apply Morrey's theorem \cite[Thm.~9.12]{Brezis10} to obtain that all $h\in W^{1,2}(\mathbb{R}_+)$ is 1/2-Hölder continuous:
	\begin{align*}
		|h(t)-h(t')|\leq C |t-t'|^{\frac{1}{2}}\|Dh\|_{\mathcal{L}_2(\mathbb{R}_+)},\ \forall t,t'\in \mathbb{R}_+. 
	\end{align*} 
	This implies the uniform continuity of $h$ by definition, and the  continuous extension theorem \cite[Thm. 5.4.8]{BS00} guarantees the finite value of the  limit $h(0^+)$. Second, we show  that $h\mapsto h(0^+)$ is a bounded functional on  $W^{1,2}(\mathbb{R}_+)$. Given that $h\in W^{1,2}(\mathbb{R}_+)\subset  W^{1,1}_{loc}(\mathbb{R}_+)$, we have that $h$ restricted to the bounded domain is absolutely continuous \cite[p. 207]{Brezis10}, and then the fundamental theorem of calculus \cite[Thm. 3.35]{Folland99} holds as 
	\begin{align*}
		h(0^+)= h(t) - \int_{0}^t Dh(s)\,d s,  \text{for any }0<t<\infty. 
	\end{align*}
	Hence, we can obtain that 
	\begin{align*}
		h(0^+)= \int_{0}^1	h(0^+) dt= \int_{0}^1 h(t)dt -\int_{0}^1 \int_{0}^t Dh(s)\,d s \, dt,
	\end{align*}
	where, by employing  Fubini's theorem, it holds that 
	\begin{align*}
		\resizebox{\columnwidth}{!}{$
			\begin{aligned}
				\int_{0}^1 \int_{0}^t Dh(s)\,d s \, dt= \int_{0}^1 \int_{s}^1 Dh(s) \, dt\,d s= \int_{0}^1 Dh(s) (1-s)\,d s.\end{aligned}$}
	\end{align*}
	Then we have 
	\begin{align*}
			|h(0^+)|\!\leq\! \|h\|_{\mathcal{L}_2((0,1))} \|1\|_{\mathcal{L}_2((0,1))}\!+\!  \|D h\|_{\mathcal{L}_2((0,1))} \|1\|_{\mathcal{L}_2((0,1))},
	\end{align*}
	which implies the boundedness of  $h\mapsto h(0^+)$ on  $W^{1,2}(\mathbb{R}_+)$. This completes the proof of $\mathcal{V}(\mathbb{R}_+)= W^{1,2}(\mathbb{R}_+)$. 
	
	Second, we aim to show the identity of  $\mathcal{V}(\mathbb{R}_+)$ and $\mathcal{H}_{\kappa^{\text{La}}}$. 
	Recall that $W^{m,2}(\mathcal{I}_o)$ is a  Hilbert space \cite[Thm 3.6]{AF03}. It can be shown that  $\mathcal{V}(\mathbb{R}_+)$ is a Hilbert space with the inner product:   
	\begin{multline}\label{eq:pf_lapalce_inner_prod}
		\langle g,f \rangle_{\mathcal{V}(\mathbb{R}_+)}=\int_{\mathbb{R}_+}Dg(t)Df(t) dt \\
		+\alpha^2 \int_{\mathbb{R}_+} g(t)f(t) \,dt	+  \alpha g(0^+)f(0^+).
	\end{multline}
	By the definition of the RKHS and reproducing kernel \cite{Aronszajn50, SC08, SS16}, it suffices to show that:
	\begin{itemize}
		\item[1)] the kernel section  $\kappa^{\text{La}}_{t'}(t):= \kappa^{\text{La}}(t,t')$   satisfies $\kappa^{\text{La}}_{t'}\in \mathcal{V}(\mathbb{R}_+), \ \forall t'\in \mathbb{R}_+$; 
		\item[2)] the reproducing property holds as	\begin{align*}
			g(t')=\langle  g, \kappa^{\text{La}}_{t'}\rangle_{\mathcal{V}(\mathbb{R}_+)}, \  \forall t'\in \mathbb{R}_+,\ \forall g\in \mathcal{V}(\mathbb{R}_+). 
		\end{align*}
	\end{itemize} 
	To  prove 1), we consider the expressions of $\kappa^{\text{La}}_{t'}(t)$ and its weak derivative as follows: 
	\begin{align*}
		& \kappa^{\text{La}}_{t'}(t)=\frac{1}{2\alpha} e^{-\alpha(t'-t)}\mathds{1}_{(0, t')}(t)+\frac{1}{2\alpha} e^{-\alpha(t-t')}\mathds{1}_{ (t', \infty)}(t),\\
		& D \kappa^{\text{La}}_{t'}(t)=  \frac{1}{2} e^{-\alpha(t'-t)}\mathds{1}_{(0, t')}(t) -\frac{1}{2} e^{-\alpha(t-t')}\mathds{1}_{ (t', \infty)}(t),
	\end{align*}
	and thus we have  $\kappa^{\text{La}}_{t'}\in \mathcal{V}(\mathbb{R}_+)$ for all $t'\in \mathbb{R}_+$.  As for the reproducing property 2), it can be shown that 
	\begin{align*}
		\resizebox{\columnwidth}{!}{$
			\begin{aligned}
				&\langle  g, \kappa^{\text{La}}_{t'}\rangle_{\mathcal{V}(\mathbb{R}_+)}= \frac{1}{2}\int_{0}^{t'}Dg(t)  e^{-\alpha(t'-t)} dt-\frac{1}{2}\int_{t'}^\infty Dg(t)  e^{-\alpha(t-t')}dt \\
				&\!+\!\frac{\alpha}{2}\int_{0}^{t'}g(t)e^{-\alpha(t'-t)} dt\!+\!\frac{\alpha}{2}	\int_{t'}^\infty g(t) e^{-\alpha(t-t')}dt\!+\! \frac{1}{2}g(0^+) e^{-\alpha(t'-0^+)}\\
				& =\frac{1}{2}\int_{0^+}^{t'} ( g(t)(\alpha) e^{-\alpha(t'-t)}+ Dg(t) e^{-\alpha(t'-t)})  dt\\
				&\!-\!	\frac{1}{2}\int_{t'}^\infty ( g(t)(-\alpha)e^{-\alpha(t-t')}\!+\!  Dg(t)e^{-\alpha(t-t')})  dt \!+\! \frac{1}{2}g(0^+) e^{-\alpha(t'-0^+)} \\
				& =\frac{1}{2}\bigg[ g(t) e^{-\alpha(t'-t)} \bigg]_{t=0^+}^{t=t'}\!-\!\frac{1}{2}\bigg[ g(t) e^{-\alpha(t-t')} \bigg]_{t=t'}^{t=\infty}\!+\! \frac{1}{2}g(0^+) e^{-\alpha(t'-0^+)}  \\
				& = g(t').
			\end{aligned}$}
	\end{align*}
	The above reproducing property shows that  $\mathcal{V}(\mathbb{R}_+)$ is exactly  $\mathcal{H}_{\kappa^\text{La}}$, which completes the proof of Lemma \ref{le:rkhs_norm_laplace_kernel}.

	\subsection{Proof of Corollary \ref{co:rkhs_norm_dc_kernel}}

	We aim to show the derivation from  \eqref{eq:rkhs_norm_aml_kernel} to  \eqref{eq:rkhs_norm_dc_kernel} with $b(t)=\sqrt{2\alpha }e^{-\frac{\beta}{2} t}$. Since $\frac{1}{b}\in  \mathcal{C}^\infty(\mathbb{R}_+)$, we have
	$g, Dg, \frac{g}{b}, D\frac{g}{b}\in \mathcal{L}_{1,\text{loc}}(\mathbb{R}_+)$ for $g\in \mathcal{H}_{\kappa^{\text{DC}}}$.
	We apply Lemma \ref{le:prod_rule_weak_derivative} such that
	\begin{align*}
		D(\frac{g}{b})(t)&= \frac{1}{b(t)} Dg(t)+ g(t) (\frac{1}{b})^{(1)}(t), \\
		&= \sqrt{\frac{1 }{2\alpha}} e^{\beta t/2}Dg(t)+ \sqrt{\frac{1 }{2\alpha}}g(t) (\frac{\beta}{2}e^{\beta t/2}).
	\end{align*}
Substitution   into \eqref{eq:rkhs_norm_aml_kernel} gives
	\begin{align*}
		&\|g\|_{\mathcal{H}_{\kappa^{\text{DC}}}}^2=  \frac{\alpha}{2}\int_{\mathbb{R}_+} |g(t)e^{\beta t/2}|^2\,dt\\
		&+\frac{1}{2\alpha}\int_{\mathbb{R}_+}  | (Dg(t)+ \frac{\beta}{2}g(t) )e^{\beta t/2}|^2\,dt 
		+ \frac{1}{2} |g(0^+)|^2   \\
		&  =(\frac{\alpha}{2}+\frac{\beta^2}{8\alpha})\int_{\mathbb{R}_+} |g(t)|^2e^{\beta t}\,dt+\frac{1}{2\alpha}\int_{\mathbb{R}_+}   \bigg( |Dg(t)|^2\\
		&+ \frac{\beta}{2} Dg(t)g(t)+ \frac{\beta}{2} Dg(t)g(t) \bigg)e^{\beta t}\,dt 
		+ \frac{1}{2} |g(0^+)|^2.
	\end{align*}
	Further, we exploit Lemma \ref{le:prod_rule_weak_derivative2} and $g\in W^{1,2}(\mathbb{R}_+)$ such that 
	\begin{align}
		&\int_{\mathbb{R}_+} (  Dg(t)g(t)+ Dg(t)g(t))e^{\beta t}\,dt=	\int_{\mathbb{R}_+}  D(gg)(t)e^{\beta t}\,dt\nonumber\\
		&=\bigg[ g(t)g(t)e^{\beta t}\bigg]_{t=0^+}^{t=\infty} -\int_{\mathbb{R}_+}  \beta|g(t)|^2 e^{\beta t}\,dt\nonumber\\
		&= - |g(0^+)|^2-\beta \int_{\mathbb{R}_+}  |g(t)|^2 e^{\beta t}\,dt, \label{eq:pf_Dgg_equ_gt2}
	\end{align}
	where $ g(t)g(t)e^{\beta t}\rightarrow 0$ as $t\rightarrow \infty$
	because of  $g/b\in W^{1,2}(\mathbb{R}_+)$  such that $\lim_{t\rightarrow \infty}g(t)/b(t)= 0$ \cite[Coro. 8.9]{Brezis10}.  Finally,  with the proper arrangement, we  obtain 
	\begin{multline}
		\|g\|_{\mathcal{H}_{\kappa^{\text{DC}}}}^2=    (\frac{\alpha}{2}-\frac{\beta^2}{8\alpha})\int_{\mathbb{R}_+} |g(t)|^2e^{\beta t}\,dt+\\
		\frac{1}{2\alpha}\int_{\mathbb{R}_+}    |Dg(t)|^2e^{\beta t}\,dt 
		+ (\frac{1}{2}-\frac{\beta}{4\alpha}) |g(0^+)|^2.\nonumber
	\end{multline}
	This completes the derivation from  \eqref{eq:rkhs_norm_aml_kernel} to  \eqref{eq:rkhs_norm_dc_kernel}. 
	
Finally, 
	the equivalence between \eqref{eq:rkhs_norm_dc_cov} and  \eqref{eq:rkhs_norm_dc_kernel} can be proved by employing the equality \eqref{eq:pf_Dgg_equ_gt2}.

\end{document}